\documentclass[pra,preprint]{revtex4-1}
\usepackage{graphicx}
\usepackage{epstopdf}

\newcommand{\e}{{\rm e}}

\newcommand{\ep}{\nu}

\newcommand{\om}{\omega}

\newcommand{\bea}{\begin{eqnarray}}
\newcommand{\eea}{\end{eqnarray}}
\newcommand{\be}{\begin{equation}}
\newcommand{\ee}{\end{equation}}
\newcommand{\ba}{\begin{eqnarray}}
\newcommand{\ea}{\end{eqnarray}}

\newcommand{\nn}{\nonumber}
\newcommand{\la}{\label}

\begin{document}
\title{Peak-height formula for higher-order breathers of the nonlinear Schr\"odinger equation on non-uniform backgrounds}

\author{Siu A. Chin$^1$, Omar A. Ashour$^{1,2}$, Stanko N. Nikoli\'c$^{2,3}$, Milivoj R. Beli\'c$^2$ }
\affiliation{$^1$Department of Physics and Astronomy,
Texas A\&M University, College Station, TX 77843, USA}
\affiliation{$^2$Science Program, Texas A\&M University at Qatar,
P.O. Box 23874 Doha, Qatar}
\affiliation{$^3$Institute of Physics, University of Belgrade,
Pregrevica 118, 11080 Belgrade, Serbia}

\begin{abstract}

Given any background (or seed) solution of the nonlinear Schr\"odinger equation, the Darboux transformation
can be used to generate higher-order breathers with much greater peak intensities. 
In this work, we use the Darboux transformation to prove, in a unified manner and without knowing 
the analytical form of the background solution,
that the peak-height of a high-order breather is just a sum of peak-heights of first-order
breathers plus that of the background, {\it irrespective} of the specific choice of the background. 
Detailed results are verified for breathers on a cnoidal background. 
Generalizations to more extended nonlinear Schr\"odinger equations, such as the Hirota equation, are indicated.

\end{abstract}


\maketitle
\section{Introduction}

The study of high-intensity optical solitons and breathers of the cubic nonlinear
Schr\"odinger  (NLS) equation, has became a cornerstone of modern 
nonlinear physics and is of special importance in modern nonlinear photonics. 
For a comprehensive review of optical
solitons and breathers, see the work by Dudley {\it et al.} \cite{dud14}.

While the inverse scattering \cite{zak72} and the direct method \cite{hir04} have been used in the past to
study solitons of the NLS equation, much of recent advances in understanding 
breathers \cite{akh86} and rogue waves \cite{akh09,akh093} are based on using the 
Darboux transformation \cite{akh88} (DT).
Given any background (or seed) solution of the NLS equation, the Darboux transformation can be
used to generate a high-order solution on top of that background with either greater
peak intensity \cite{akh86,akh09,akh093}, or greater shape complexity \cite{ked112,ked13}. 
 
The Darboux transformation is generic in that by iterating a pair of generating solutions of the Lax-pair \cite{lax68} 
equation containing the background as an input function, it provides a systematic procedure for creating new solutions. 
The Darboux transformation itself knows nothing about the evolution equation it is transforming, 
nor the background wave function it is using.
The Darboux iterations are therefore the same for the cubic NLS equation,
the Hirota equation \cite{hir73,ank10,cho152}, and other extended NLS equations \cite{ank14,cho14,cho15},
regardless of the choice of the background. Variants in the evolution
equation and the background solution are only reflected in the initial generating solutions of the Lax-pair
equation. Different evolution equations have different Lax-pair equations and therefore 
different generating solutions. The same Lax-pair equation with different background solutions
will also have different generating solutions. 

In this work, we first prove in Section \ref{fund} a remarkable generic result for DT. 
If the initial generating functions
of DT satisfy a simple phase condition (\ref{fkey}) below, then the peak-height formula (\ref{peak})
follows. This formula states that the peak-height of a high-order soliton/breather is 
just a sum of its constituent first-order soliton/breather peak-heights plus that of the background. 
Thus, as long as (\ref{fkey}) is true, regardless of the choice of evolution equation or
the background wave function, one has the peak-height formula (\ref{peak}). Hence, Eq. (\ref{fkey})
guarrantees the peak-height formula for the cubic NLS equation, the Hirota equation \cite{hir73,ank10,cho152}, 
and any other extended NLS equation \cite{ank14, cho14, cho15} that evolves according to the
Lax-pair equation \cite{lax68}. This then greatly generalizes the peak-height formula first stated 
for solitons \cite{akh91} and more recently proved for Akhmediev breathers \cite{chin16}. 
We recall that in Ref. \cite{chin16}, we have shown that the peak-height formula on a constant
background is essential for determining what first-order breathers are necessary for producing
a higher-order breather of a given intensity. One can then extract an initial profile of the light pulse, 
with the correct Fourier components, so that when such a pulse is initiated in an optical fiber 
(assuming that its propagation is well-described by the NLS equation), will be compressed into 
a breather of the required intensity. Such an initial light pulse can be produced in experiments 
similar to those described in Refs. \cite{erk11,fri14}, particularly via the latter reference's
frequency-comb. By proving a more general peak-height formula here, we hope to pave the 
way for a possible future practical realization of these more general NLS equations. 

The phase condition (\ref{fkey}) however, is simply a relative phase between the two
generating functions of the Lax equation, and can always be conveniently so chosen 
in the soliton case of $\psi_0=0$. This is also shown in Section \ref{fund}.
Thus, the peak-height formula holds for solitons of all the NLS equations mentioned above.

Furthermore, when $\psi_0\ne 0$, the background generates a non-trival phase for the two generating
functions of the Lax equation. We prove in Section \ref{back} that for the cubic NLS equation,
and without knowing the analytical form of the background wave function, Eq.
(\ref{fkey}) remains true despite the added background phase. That is, the peak-height formula 
(\ref{peak}) for the cubic NLS equation is true regardless of the choice of the background solution: vanishing, 
uniform, or varying. In this manner, the proof of the peak-height formula for the 
cubic NLS equation is made complete. 

For the Hirota and other extended NLS equations on a uniform background, 
others \cite{hir73,ank10,cho152,ank14, cho14, cho15} have shown that
(\ref{fkey}) is true and therefore the peak-height formula also holds. However,
for a non-uniform background, the Lax-pair equations for these extended equations
are more complex and it is difficult and  beyond the scope of this work   
to prove (\ref{fkey}) for the extended equations on a general background.

Finally, in Section \ref{cno} we show that for the cubic NLS equation, 
the only non-uniform background that can support
Akhmediev-type breathers is the Jacobi elliptic function dn$(t,k)$, which forms a ``dnoidal" background.
In the end, in Section \ref{num} we verify our theoretical results with numerical calculations and summarize
our conclusions in Section \ref{con}.

\section{Basic result for the Darboux transformation}
\la{fund}

The $N$th-order DT wave function of any nonlinear evolution equation,
such as the cubic nonlinear Schr\"odinger equation
\be
i\frac{\partial\psi}{\partial x}
+\frac12 \frac{\partial^2\psi}{\partial t^2}+|\psi|^2\psi=0,
\la{sch}
\ee
is given by \cite{akh88}
\be
\psi_N(x,t)=\psi_0(x,t)+\sum_{n=1}^N\frac{2(l^*_n-l_n)s_{n1}r_{n1}^*}{|r_{n1}|^2+|s_{n1}|^2},
\la{dtwf}
\ee
where $\psi_0(x,t)$ is the background solution and the sum goes over $N$ constituent 
soliton/breather solutions, each characterized by an eigenvalue 
$$
l_n=i\ep_n, \quad{\rm with}\quad \ep_n>0.
$$
Here, $x$ and $t$ are the conventional propagation distance and transverse variable of fiber optics. 
At a given $n$, the functions $r_{n1}(x,t)$ and $s_{n1}(x,t)$ depend recursively on all the
lower $n$ functions via \cite{akh88}
\ba
r_{nj}=[&&  (l_{n-1}^*-l_{n-1})s^*_{n-1,1}r_{n-1,1}s_{n-1,j+1}\nn\\
           &&+(l_{j+n-1}-l_{n-1})|r_{n-1,1}|^2r_{n-1,j+1}\nn\\
          && +(l_{j+n-1}-l^*_{n-1})|s_{n-1,1}|^2r_{n-1,j+1}]/(|r_{n-1,1}|^2+|s_{n-1,1}|^2),\la{eqr}\\
s_{nj}=[&&  (l_{n-1}^*-l_{n-1})s_{n-1,1}r^*_{n-1,1}r_{n-1,j+1}\nn\\
           &&+(l_{j+n-1}-l_{n-1})|s_{n-1,1}|^2s_{n-1,j+1}\nn\\
           &&+(l_{j+n-1}-l^*_{n-1})|r_{n-1,1}|^2s_{n-1,j+1}]/(|r_{n-1,1}|^2+|s_{n-1,1}|^2).
\la{eqs}
\ea
These DT iterations are ``generic" in that they are of the same form for all NLS or extended equations 
they are designed to solve. The knowledge of a particular nonlinear equation or a background
solution is encoded only in the initial solutions 
$
r_{1j}(x,t)\quad{\rm and}\quad s_{1j}(x,t)
$
of the Lax-pair equation (to be described below), 
which kick-start the iterations of (\ref{eqr}) and (\ref{eqs}).

For the general wave function $\psi_N(x,t)$, iterations (\ref{eqr}) and (\ref{eqs}) are
recursively too complex to be written down analytically beyond the lowest few orders. 
However, we can prove a fundamental result on the
basis of (\ref{eqr}) and (\ref{eqs}) alone, that if for all $1\le n \le N$,
$s_{n1}(0,0)$ and $r_{n1}(0,0)$ only differ by an arbitrary phase $\phi$,
\be
s_{n1}(0,0)=\e^{i\phi}  r_{n1}(0,0),
\la{key}
\ee
then
\be
\psi_N(0,0)=\psi_0(0,0)+\sum_{n=1}^N(-i\e^{i\phi}) 2\ep_n=\psi_0(0,0)+\sum_{n=1}^N 2\ep_n.
\la{peak}
\ee
This {\it peak-height formula} (\ref{peak}) follows from (\ref{key}) by simply evaluating (\ref{dtwf}) at the
origin $x=t=0$ with the choice of the phase 
$$
\phi=\frac{\pi}{2}.
$$
Note that (\ref{peak}) gives the wave function itself, not its modulus. Since one can always 
center the soliton/breather at the origin, this formula gives the peak-height of the
$N$th-order soliton/breather as a {\it linear} sum of peak-heights of individual soliton/breathers
plus the height of the background solution. The choice of the phase 
$\e^{i\phi}=i$ is natural, in that the resulting peak-height is real and positive. 

We will now prove that (\ref{key}) is true if the initial Lax solutions also satisfy the phase condition
\be
s_{1j}(0,0)=\e^{i\phi} r_{1j}(0,0),
\la{fkey}
\ee
for each $j$-constituent soliton/breather. We defer the proof of (\ref{fkey}) to the next section,
since these initial Lax solutions require knowledge of the specific equation and the background.

To prove (\ref{key}) on the basis of (\ref{fkey}), 
we apply iterations (\ref{eqr}) and (\ref{eqs}) at $x=t=0$ and suppress the notation $(0,0)$. 
Starting from (\ref{fkey}), which is $s_{1j}=\e^{i\phi}r_{1j}$ for $1\!\le\! j\!\le\! N$, one
can prove successively that $s_{2j}=\e^{i\phi}r_{2j}$ for $1\!\le\! j\!\le\! N\!-\!1$,
$s_{3j}=\e^{i\phi}r_{3j}$ for $1\!\le\! j\!\le\! N\!-\!2$, {\it etc.}. Therefore,
given $s_{n-1,j}=\e^{i\phi}r_{n-1,j}$, (\ref{eqr}) and (\ref{eqs}) reads
\ba
r_{nj}
&=&-(i\e^{-i\phi})\ep_{n-1}\,s_{n-1,j+1}+i\ep_{j+n-1}\,r_{n-1,j+1},\nn\\
s_{nj}
&=&(-i\e^{i\phi})\ep_{n-1}\,r_{n-1,j+1}+i\ep_{j+n-1}\,s_{n-1,j+1},\nn
\ea
which means that
\ba
\e^{i\phi}r_{nj}&=&-i\ep_{n-1} s_{n-1,j+1}+i\ep_{j+n-1} (\e^{i\phi} r_{n-1,j+1})\nn\\
&=&(-i\e^{i\phi})\ep_{n-1}\,r_{n-1,j+1}+i\ep_{j+n-1}\,s_{n-1,j+1}\nn\\
&=&s_{nj}.\nn
\ea
Thus, the proof by induction is complete. Note that we only need the above equality
to evaluate (\ref{dtwf}) for the peak-height; 
we {\it do not} need the actual analytical expression for $r_{nj}$ and $s_{nj}$.
This is why the peak-height can be evaluated simply, circumventing the full 
nonlinear complexity of DT.

In the case of the cubic NLS equation, the initial functions
$s_{1j}(x,t)$, $r_{1j}(x,t)$ are solutions to the following set of four Lax-pair partial differential equations \cite{akh09} 
(subscripts dropped for clarity)
\ba
r_t&=&i\psi_0^*s+ilr\la{eqrt},\\
s_t&=&i\psi_0r-ils\la{eqst},\\
r_x&=&(il^2-\frac{i}{2}|\psi_0|^2)r+(il\psi_0^*+\frac12(\psi_0^*)_t) s \la{eqrx},\\
s_x&=&(-il^2+\frac{i}{2}|\psi_0|^2)s+(il\psi_0-\frac12(\psi_0)_t) r, 
\la{eqsx}
\ea 
whose compatibility condition requires $\psi_0$ to be a solution of the NLS equation.
For more extended NLS equations, the Lax-pair equations remain linear,
but are more complex, and can be found in Ref. \cite{ank14, cho14, cho15}. 
Since (\ref{eqrt})-(\ref{eqsx})
is a set of four {\it linear} equations, the solutions $s_{1j}(x,t)$ and $r_{1j}(x,t)$ 
will contain four constants of integration. As the DT wave function (\ref{dtwf})
is unaffected by a common phase (or a common scale factor) of $s_{1j}(x,t)$ and $r_{1j}(x,t)$,
one constant can be used to  normalize both to unit modulus. Two constants can be
used to shift the solution peak to $x_0$ and $t_0$ and the last constant can be chosen to 
fix the relative phase, so that (\ref{fkey}) is true. Such a relative phase then guarantees
a positive wave function peak-height. This argument suggests that (\ref{fkey})
is always possible (when $x_0\!=\!t_0\!=\!0$) by simply choosing
\be
r_{1j}(0,0)=\e^{-i\pi/4} \quad{\rm and}\quad s_{1j}(0,0)=\e^{i\phi}r_{1j}(0,0)=\e^{i\pi/4}.
\la{easy}
\ee
This is indeed the case for solitons based on the background $\psi_0=0$. 
For example, integrating (\ref{eqrt})-(\ref{eqsx}) with $\psi_0=0$ gives
     \ba
          r_{1j}(x,t) &=& \exp[-\ep_j (t-t_0) - i\ep_j^2 (x-x_0)-i\pi/4], \la{rs0}\\
          s_{1j}(x,t) &=& \exp[\ \ \ \ep_j (t-t_0) + i\ep_j^2 (x-x_0)+i\pi/4],
\la{rs1}
     \ea
which produces a first-order soliton (when setting $x_0\!=\!t_0\!=\!0$) 
     \ba
     \psi_1(x,t) &=& \psi_0 + \frac{2(l_1^* - l_1)r_{11}^*s_{11}}{|r_{11}|^2 + |s_{11}|^2}
                    = \frac{2(-2i\ep_1)\exp(i2\ep_1^2x+i\pi/2)}{\exp(-2\ep_1 t) + \exp(2\ep_1 t)}\nn\\
                    &=& \frac{2\ep_1\exp(i2\ep_1^2x)}{\cosh(2\ep_1 t)}
\nn
     \ea
with a positive peak-height at the origin  
\be
\psi_1(0,0)=2\ep_1.
\ee
Since (\ref{rs0}) and (\ref{rs1}) satisfy (\ref{fkey}) when $x_0\!=\!t_0\!=\!0$, 
the peak-height of an $N$th-order soliton is given 
by (\ref{peak}), with $\psi_0(0,0)=0$. This result has been stated without a proof  some time ago \cite{akh91},
but with a different relative phase between $r_{1j}(x,t)$ and $s_{1j}(x,t)$. Result (\ref{easy}) remains
true for $\psi_0=0$ in the more extended NLS equations, including the Hirota and
the Lakshmanan-Porsezian-Daniel operators, see Ref. \cite{ank14}. 
Thus, the peak height formula holds for solitons in all extended NLS equations.

When $\psi_0\ne 0$, as we shall see below, the background will generate an additional phase on the initial Lax solutions, and one must show that (\ref{fkey}) remains true in spite of  the added phase.

\section{General background wave functions}
\la{back}

For solutions of the NLS equation with a non-uniform background of the form
\be
\psi_0(x,t)=AF(t)\e^{iBx},
\la{gens}
\ee
where $A\ne 0$ and $F(t)$ is real, we will show that Eq.  (\ref{fkey}) remains true,
but now {\it requires} that $\phi=\pi/2$.
For $\psi_0(0,0)$ to be the peak, we assume that $F(t)$ is
normalized such that 
\be
F(0)=1\qquad{\rm and}\qquad F_t(0)=0.
\la{itc}
\ee
This is all that we need to prove (\ref{fkey}); we do not need to know the analytic form of $F(t)$.
(We also do not need to require $F_{tt}(0)<0$; the case of $F_{tt}(0)>0$ is covered by taking $A<0$.)

For $F(t)$ not constant, one cannot solve all four Lax-pair equations. 
However, one can still solve (\ref{eqrx}) and (\ref{eqsx}) by invoking (\ref{gens}) and (\ref{itc}). 
Fixing $t=0$ with $l=i\ep$, the last two Lax-pair equations read  
(suppressing the subscripts and $(x,0)$ dependence)
\ba
r_x&=&-i(\ep^2+\frac{1}{2}A^2)r-\ep A\e^{-iBx}  s \nn\\
s_x&=&\ \ i(\ep^2+\frac{1}{2}A^2)s-\ep A\e^{iBx}r. 
\nn
\ea 
Letting
$$
r=a\e^{-iBx/2}\qquad{\rm and}\qquad s=b\e^{iBx/2}
$$
gives
\ba
a_x&=&-iUa-\ep A b ,\la{eqax}\\
b_x&=&\ \ iUb-\ep A a ,\la{eqbx} 
\ea
where $U=\ep^2+\frac{1}{2}(A^2-B)$. It follows that
\be
b_{xx}=(\ep^2A^2-U^2)b=\ep^2\om^2b
\la{bb}
\ee
with
\be
\om=\sqrt{B-\ep^2-(\frac{A^2-B}{2\ep})^2},
\ee
and hence the solution is
$$
b=C\e^{\ep\om x}+D\e^{-\ep\om x}.
$$
In this work, we focus on the breathers 
with real $\om$, thus restricting
\be
B-\ep^2-\frac1{\ep^2} (\frac{A^2-B}{2})^2\ge 0.
\la{res}
\ee
The boundary value of this equation defines the rogue wave limit of $\om=0$.
When $\om$ is imaginary, one
has Kuznetsov-Ma-type solutions periodic in $x$  \cite{kuz77,ma79}.

For breathers, the solution for $a$ follows from (\ref{eqbx}),
$$
a=\frac1{\ep A}(i Ub-b_x)=D\left(\frac{\ep\om+iU}{\ep A}\right)\e^{-\ep\om x}
-C\left(\frac{\ep\om-iU}{\ep A}\right)\e^{\ep\om x}.
$$
From (\ref{bb}), the parentheses are just pure phases,  
$$
a=D\e^{i 2\chi -\ep\om x}-C\e^{-i 2\chi+\ep\om x},
$$
given by
\be
\cos(2\chi)=\om/A\qquad{\rm and}\qquad \sin(2\chi)=U/(\ep A).
\la{chi}
\ee
To make $a$ and $b$ symmetrical, we can chose 
$$
D=\e^{-i\chi-i\delta}\qquad{\rm and}\qquad C=\e^{i\chi+i\gamma},
$$
with phases $\delta$ and $\gamma$ yet to be determined.
This then gives, restoring all subscripts,
\ba
r_{1j}(0,0)&=&\e^{i\chi_j-i\delta}-\e^{-i\chi_j+i\gamma}=\e^{i\chi_j-i\delta}+\e^{-i\chi_j+i\gamma-i\pi},        \nn\\
s_{1j}(0,0)&=& \e^{i\chi_j+i\gamma}+\e^{-i\chi_j-i\delta}.\nn
\ea
The above will satisfy (\ref{fkey}) if the phase $\phi$ is given by 
$$
\phi=\delta+\gamma=\frac{\pi}2 .
$$
Note that only the sum $\delta+\gamma$ is fixed to be $\pi/2$, however, the symmetrical
choice of $\delta=\gamma=\pi/4$ is universally adopted in the literature \cite{akh093,akh88,ked14}.
Thus, for the above choice of $\phi$, (\ref{fkey}) remains true
{\it independent of the phase} $\chi_j$ {\it generated by the background}.
Therefore, the peak-height formula (\ref{peak}) for breathers is true regardless
of the choice of the background.

To verify that $2\ep_1$ is indeed the peak-height of the first-order DT wave function, we compute
directly
\ba
\psi_1(x,0)&=&\psi_0(x,0)+\frac{2(l^*_1-l_1)s_{1j}r_{11}^*}{|r_{11}|^2+|s_{11}|^2}\nn\\
&=&\biggl(A+\frac{2\ep_1[1-\sin(2\chi_1)\cosh(2\ep_1\om_1 x)+i\cos(2\chi_1)\sinh(2\ep_1\om_1 x)]}
{\cosh(2\ep_1\om_1 x)-\sin(2\chi_1)}\biggr)\e^{iBx}
\la{abx}
\ea 
At $x\rightarrow\pm\infty$, we have
intensity 
\ba
|\psi_1(\pm\infty,0)|^2&=&[A-2\ep_1\sin(2\chi_1)]^2+[2\ep_1\cos(2\chi_1)]^2\nn\\
&=&A^2-4A\ep_1\sin(2\chi_1)+4\ep_1^2=A^2-4U+4\ep_1^2\nn\\
&=&2B-A^2,
\la{bgd}
\ea 
{\it provided that} $\om\ne 0$.
For Akhmediev breathers on a uniform background $\psi_0=\e^{ix}$ with
$A=B=1$, the above reproduces the background intensity
$|\psi_1(x\rightarrow\pm\infty,0)|^2=1$, as compared to the peak-height $\psi_1(0,0)=1+2\ep_1$.

Breathers of the Hirota and extended NLS equations on the uniform background 
have initial Lax solutions satisfying (\ref{fkey}) \cite{cho152,cho15}. Therefore, these breathers obey the same 
peak-height formula as the Akhmediev breathers.

\section{Cnoidal background breathers}
\la{cno}

For solutions of the cubic NLS equation with non-uniform backgrounds, we
substitute (\ref{gens}) into (\ref{sch}), to find
$$
\frac{d^2 F}{dt^2}=2BF-2A^2F^3.
$$
Comparing this to the equation satisfied by any of the 12 Jacobi elliptic functions \cite{sch15} $zn(t,k)$
$$
\frac{d^2 zn}{dt^2}=\beta zn+2\alpha zn^3,
$$
we must have $\alpha=-A^2$ and $\beta=2B$. Among the 12 elliptic functions, only four have $\alpha$ negative \cite{sch15}, given by
\ba
1)\ F(t)&=&{\rm cn}(t,k)\quad {\rm with}\quad A=k, B=k^2-1/2,
\nn\\
2)\ F(t)&=&{\rm dn}(t,k)\quad {\rm with}\quad A=1, B=1-k^2/2,
\nn\\
3)\ F(t)&=&{\rm nd}(t,k)\quad {\rm with}\quad A=\sqrt{1-k^2}, B=1-k^2/2,
\nn\\
4)\ F(t)&=&{\rm sd}(t,k)\quad {\rm with}\quad A=k\sqrt{1-k^2}, B=k^2-1/2,
\nn
\ea
where $k$ is the modulus of the elliptic function. For cases 1) and 2), $t$=0  
is the peak of cn$(t,k)$ and dn$(t,k)$. However, for cases 3) and 4) the peaks are at nd$(K,k)$ and
sd$(K,k)$, where $K$ is the quarter-period of cn. Therefore, for these to peak at $t=0$, 
one must set $F(t)$=nd$(t+K,k)$ and
$F(t)$=sd$(t+K,k)$. For case 3) this means
$$
AF(t)=\sqrt{1-k^2}\, {\rm nd}(t+K,k)=\frac{\sqrt{1-k^2}}{{\rm dn}(t+K,k)}={\rm dn}(t,k),
$$
which is identical to case 2). For case 4), one has
\ba
AF(t)&=&k\sqrt{1-k^2}\, {\rm sd}(t+K,k)=k\sqrt{1-k^2}\, \frac{{\rm sn}(t+K,k)}{{\rm dn}(t+K,k)}\nn\\
&=&k\,{\rm dn}(t,k)\, \frac{{\rm cn}(t,k)}{{\rm dn}(t,k)}=k\,{\rm cn}(t,k),
\nn
\ea
which is identical to case 1). There are therefore only two cnoidal background solutions, 1) and 2). 
For these two cases respectively, our general formula (\ref{chi}) for the background phase,
\ba
\cos(2\chi_j)&=&\sqrt{1-\frac1{k^2}\Bigl(\ep_j+\frac1{4\ep_j}\Bigr)^2},\nn\\
\cos(2\chi_j)&=&\sqrt{1-\Bigl(\ep_j+\frac{k^2}{4\ep_j}\Bigr)^2 },
\nn
\ea
agrees with the results of Ref. \cite{ked14}. 

\begin{figure}
		\includegraphics[width=0.95\linewidth]{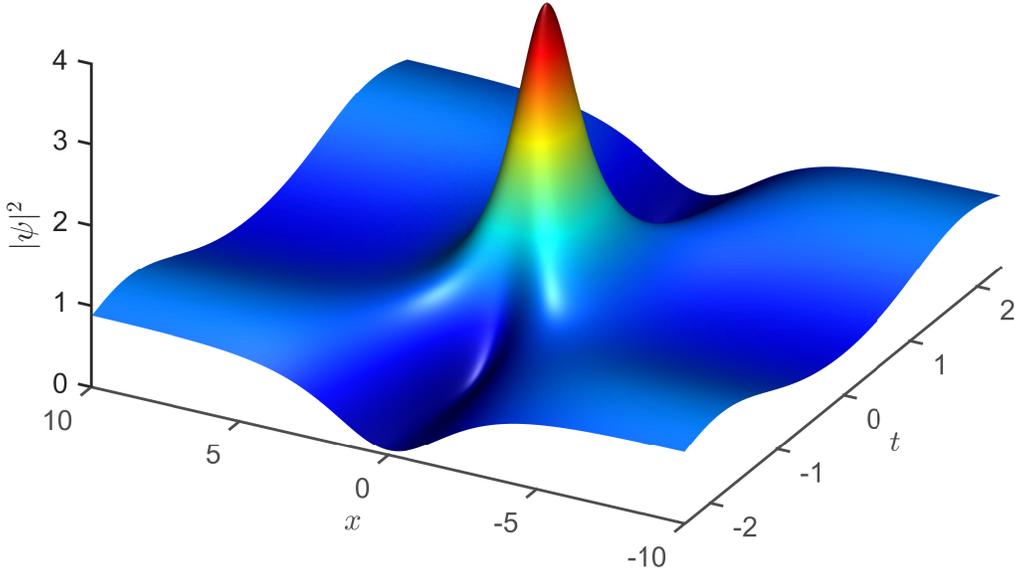}
       \caption{A breather on the dn$(t)$ elliptic function background, with $\ep=1/2$ and $k^2=1/2$.
       The peak intensity is 4.
} 
\la{hh}
\end{figure}

For the cn background, the restriction (\ref{res}) for breathers gives
$$
k^2\ge(\ep+\frac1{4\ep})^2\ge 1 ,
$$
since the RHS has a minimum of 1. There is therefore no breather except possibly at $k=1$. 
However, at $k\!=\!1$, cn$(t,1)$=sech$(t)$, which is the same as the dn case at $k=1$, described
below. Thus, all breathers are contained in the dn case and the cn background only
supports the Kuznetsov-Ma-type \cite{kuz77,ma79} breathers.

\begin{figure}
		\includegraphics[width=0.95\linewidth]{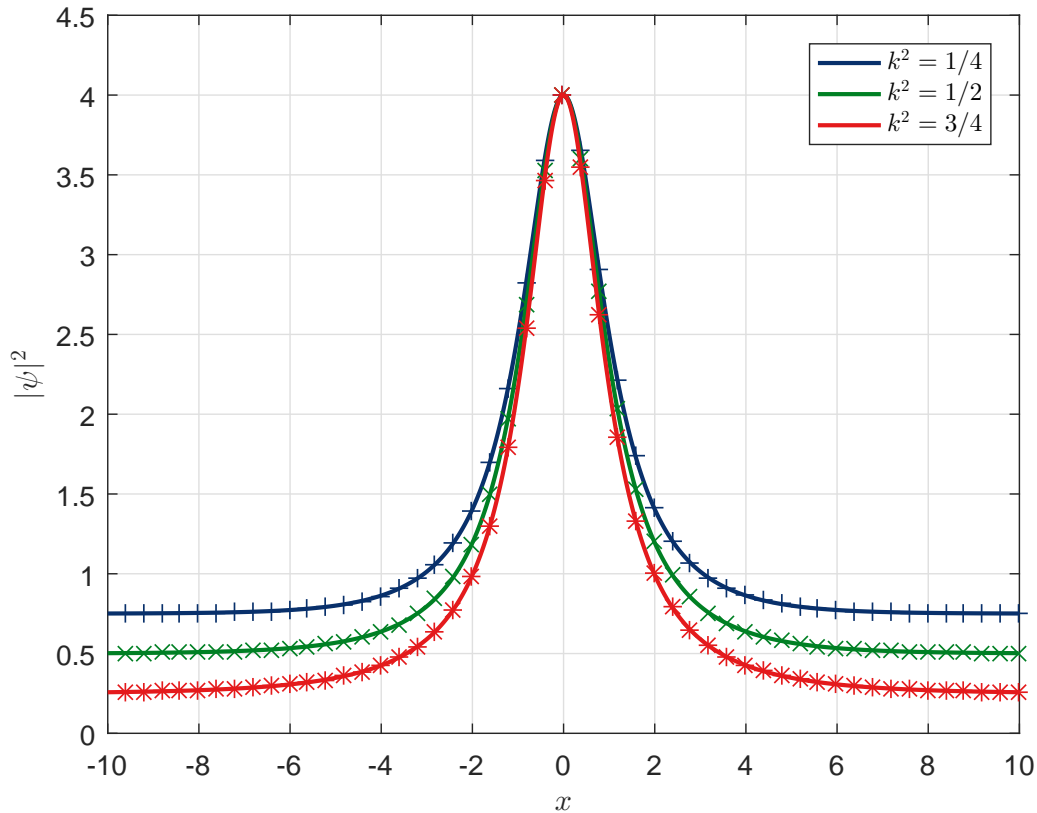}
       \caption{Comparing the DT evolution of $|\psi(x,0)|^2$ (symbols) with prediction (\ref{abx}) (lines) at
$\ep=1/2$ but at three values of $k^2=1/4,1/2,3/4$. The peak intensity at the origin is 4 for
all three cases, but the background intensities given by (\ref{dnbk}) are
respectively 3/4, 1/2 and 1/4. 
} 
\la{xevol}
\end{figure}

For the dn background, the breather condition (\ref{res}) gives
$$
(\ep+\frac{k^2}{4\ep})^2\le 1\quad\rightarrow\quad k^2\le 4\ep(1-\ep),
$$
which restricts the range of $\ep$ to
\be
\frac12-\frac12\sqrt{1-k^2}\le\ep\le \frac12+\frac12\sqrt{1-k^2}.
\la{vran}
\ee
As $k$ ranges from 0 to 1, the range of $\ep$ narrows from [0,1] to [1/2,1/2].
In contrast to the Akhmediev breather case, where there is only a single rogue
wave at $\ep=1$, corresponding to the Peregrine breather, here, at each value of $k$, there are {\it two}
rogue waves, corresponding to the lower and upper boundary values of (\ref{vran}). 
(These are called the DCRW and CCRW respectively in Ref. \cite{ked14}). The 
peak intensity of the brighter rogue wave is 
\be
|\psi_1(0,0)|^2=(2+\sqrt{1-k^2})^2,
\la{dnpk}
\ee
to be compared to the background intensity from (\ref{bgd}):
\be
|\psi_1(x\rightarrow\pm\infty,0)|^2=1-k^2.
\la{dnbk}
\ee 
While the absolute peak intensity (\ref{dnpk}) is less than that of the Peregrine breather,
its ratio with respect to the background intensity (\ref{dnbk}) is always greater than 9, with increasing $k$.

\begin{figure}
		\includegraphics[width=0.95\linewidth]{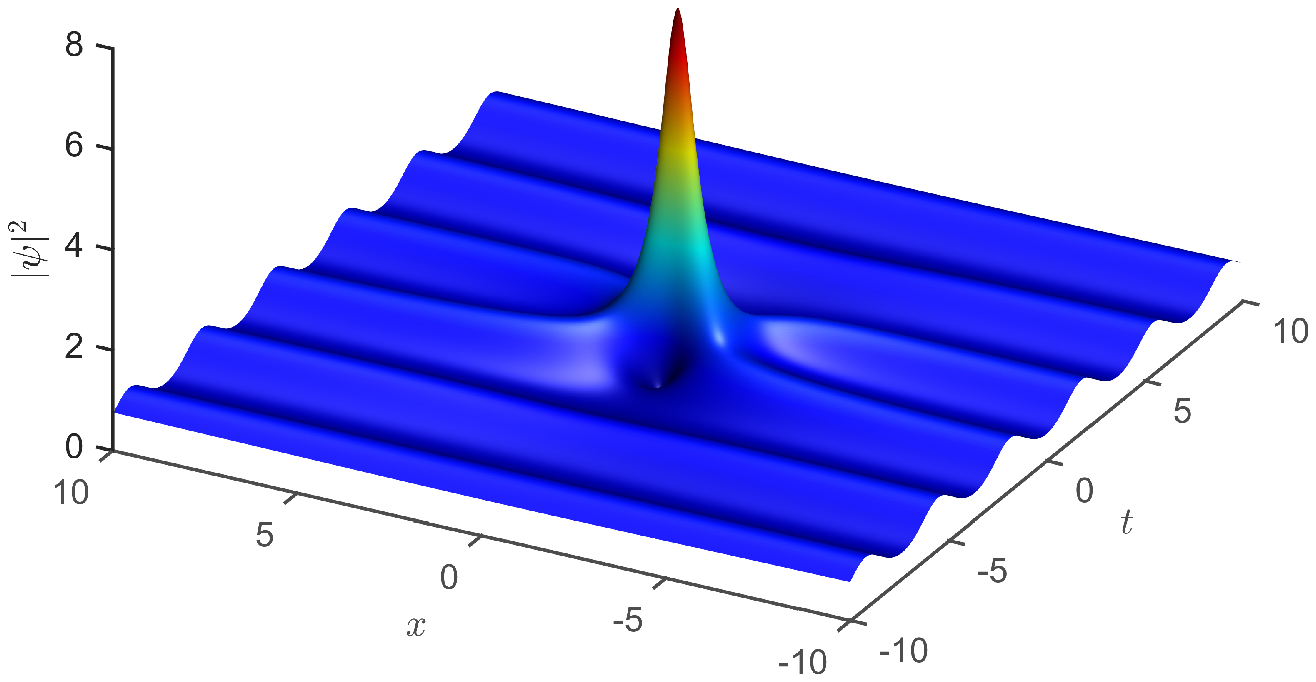}
       \caption{The bright rogue wave at $k^2=1/2$ and $\ep=\frac12+\frac12\sqrt{1-k^2}$.
The numerical peak intensity is 7.3284; the peak-height formula intensity is $4+1/2+2\sqrt{2}= 7.3284$.
} 
\la{drw}
\end{figure}

\begin{figure}
		\includegraphics[width=0.95\linewidth]{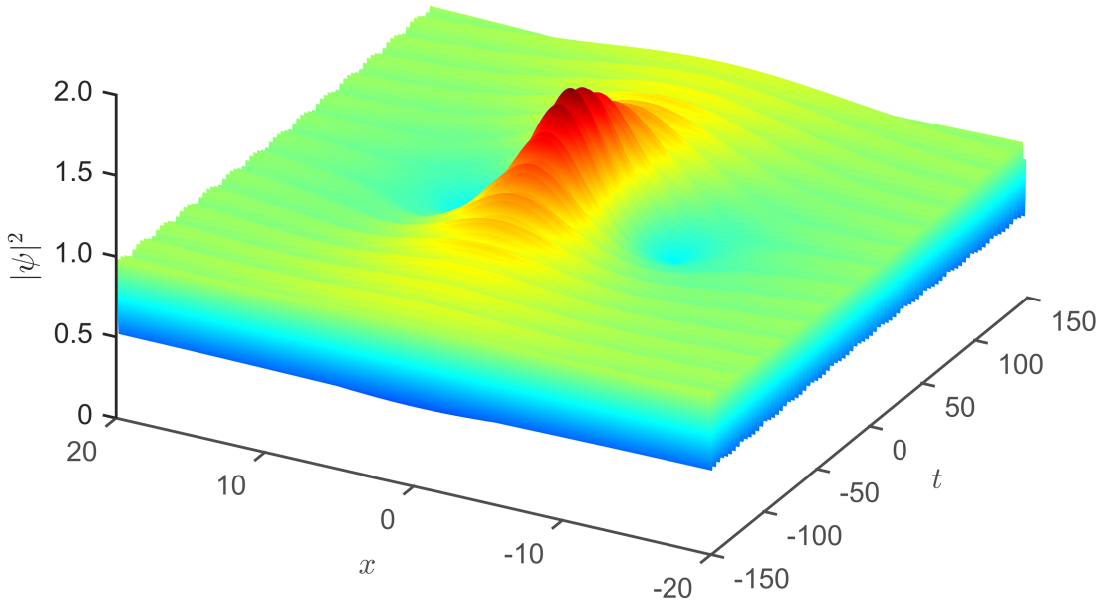}
       \caption{The dim rogue wave at $k^2=1/2$ and $\ep=\frac12-\frac12\sqrt{1-k^2}$.
The numerical peak intensity is 1.6716; the peak-height formula intensity is $4+1/2-2\sqrt{2}=1.6716$.
} 
\la{ddrw}
\end{figure}

\begin{figure}
	\includegraphics[width=0.95\linewidth]{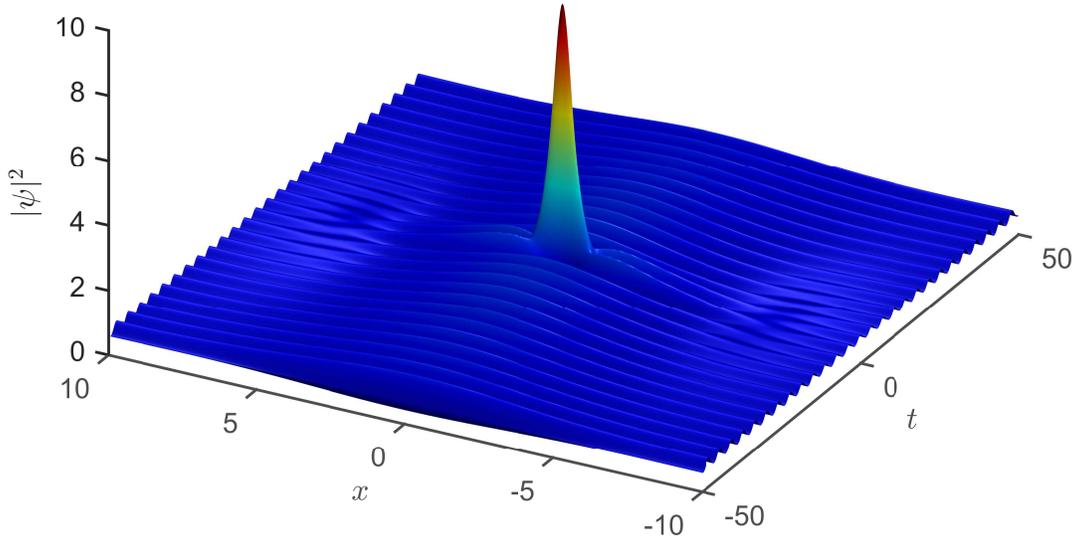}
	\caption{Second-order rogue wave at $k^2=1/2$ formed by the bright and dim rogue waves of Fig. \ref{drw} and Fig. \ref{ddrw}. The peak intensity of 9 here is precisely the sum of those two figures' intensities.} 
	\la{rw2}
\end{figure}

\begin{figure}
	\includegraphics[width=0.95\linewidth]{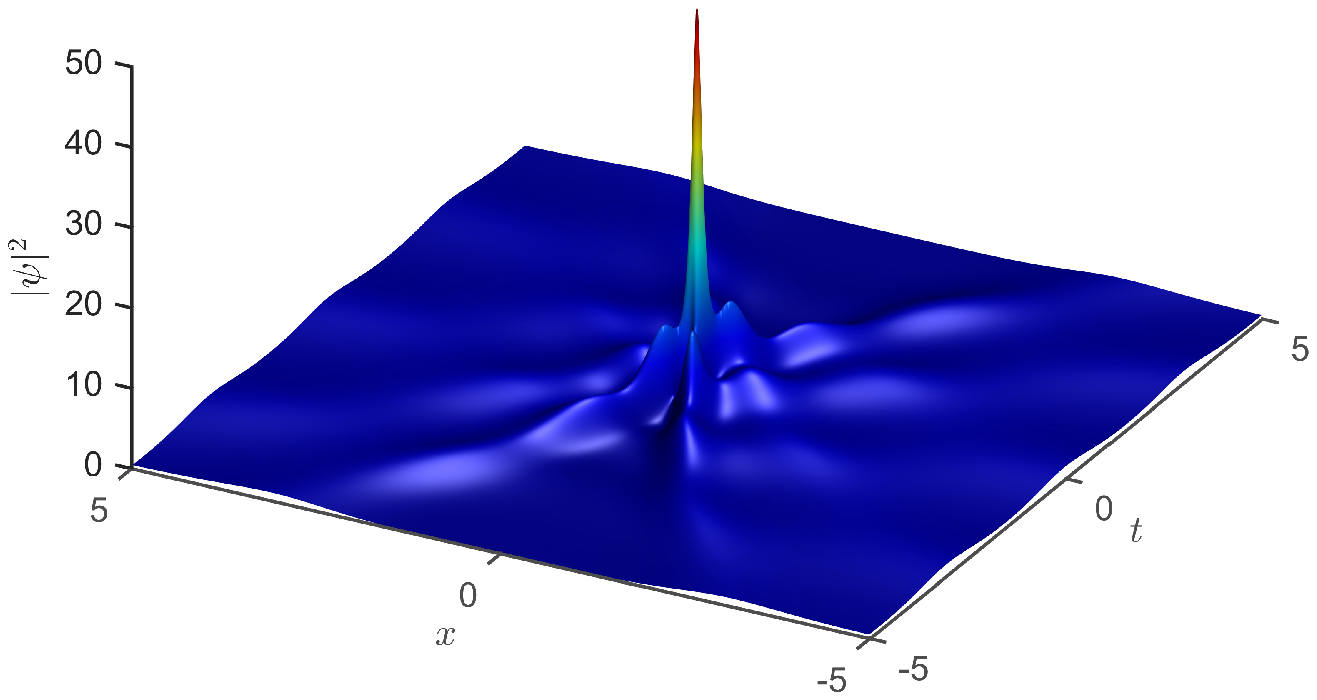}
	\caption{Fifth-order breather at $k^2=1/2$, with five values of $\ep_i=0.4,0.5,0.6,0.7,0.8$. 
		The numerical DT peak intensity is 49. The peak-height formula intensity is also 49.
	} 
	\la{fif}
\end{figure}

\section{Numerical Verification}
\la{num}

We have implemented the Darboux transformation on cnoidal backgrounds with the same initial
conditions as used by Kedziora {\it et al.} \cite{ked14}.

In Fig. {\ref{hh}, we show a breather on the dn$(t)$ elliptic function background at $\ep=1/2$ and $k^2=1/2$.
The peak intensity is precisely 4, in agreement with the peak-height formula. While the evolution
of $|\psi(x,0)|^2$ given in Fig. \ref{xevol} only reaches a single peak, its variation in the $t$-direction
is modulated by the periodic elliptic function background.

In Fig. \ref{xevol}, the DT profile $|\psi(x,0)|^2$ is compared to the general theoretical result (\ref{abx})
at $\ep=1/2$, but at three values of $k^2=1/4,1/2,3/4$. Since the peak-height only depends on $\ep$,
all three cases have the same peak intensity of 4. However, the background intensity changes according to
$k^2$, as given by (\ref{dnbk}). The agreement between numerical DT simulations and theoretical
predictions is perfect.

For the case of $k^2=1/2$, the bright rogue wave is at $\ep=\frac12+\frac12\sqrt{1-k^2}$,
with peak intensity $(1+2\ep)^2= 7.3284$. This is shown in Fig. \ref{drw}. Rogue waves on a uniform
background reaches its peak monotonically in both the $x$ and $t$ directions. Rogue waves on a cnoidal background
arises out of a background periodic in the $t$-direction.

In Fig. \ref{ddrw}, we show the dim rogue wave with $\ep=\frac12-\frac12\sqrt{1-k^2}$
at the same value of $k^2=1/2$. The intensity is more than a factor of four dimmer and with much shorter wavelength
oscillations in the $t$-direction, due to the smaller $\ep$. This intensity profile is similar to Fig. 8(a) of Ref. \cite{ked14}. 

Using DT, one can form a second-order rogue wave by utilizing these two values of the bright and dim rogue waves.
This rogue wave is special in that, since the plus and minus terms
in $\ep$ cancel, its intensity is always 9 (by our formula) {\it independent} of the background parametrized by $k$.
Its intensity is exactly at the border between first and second-order breathers.
This is shown in Fig. \ref{rw2}. Such a rogue wave was originally suggested, but not computed, 
by Kedziora {\it et al.} \cite{ked14}.
What is even more remarkable is that the intensity of this second-order rogue wave
 is exactly the sum of the intensities of the previous two first-order rogue waves! 
We have therefore found a ``Pythagorean triplet" of rogue waves. Such a result would seem 
inexplicable and mysterious in the realm of nonlinear phenomena, had it not been for the theory presented here. The power of our peak-height formula is that, in order for
$$
(2-\sqrt{1-k^2})^2+(2+\sqrt{1-k^2})^2=9
$$ 
the two first-order rogue waves must reside on a background of $k^2=1/2$! 
Thus our peak-height formula makes it easy to see that 
among all possible rogue waves given by (\ref{vran}), this Pythagorean triplet 
is unique to the background of $k^2=1/2$. There are no such rogue waves triplets on other backgrounds. 
 
A fifth-order breather is shown in Fig. \ref{fif}. Such a high-order breather is very concentrated and one
has to zero-in on the origin, to see the extremely high, yet narrow, peak. Our peak-height formula perfectly
predicted the peak-height of this high-order breather.

\section{Conclusions}
\la{con}
In this work, we have shown that for the NLS equation, the peak-height formula (\ref{peak}) is true
for all proper choices of the background solution---vanishing, uniform, or varying. 

More generally, we have also shown that,
since the DT iterations are generic, as long as (\ref{fkey}) is true, the peak-height formula (\ref{peak}) is 
true for all extended NLS equations. Such a peak-height formula will be useful in 
guiding the design and production of maximal-intensity breathers in physical systems that can be modelled 
by the NLS equation \cite{chin16} and its extended variants. Also, while there is 
no direct generalization of DT to higher spatial dimensions for the NLS equation,
the 1D solution of the NLS can be embedded into the 3D solution via 
similarity reductions \cite{yan10,zon14} in the study of Bose-Einstein condensates.
The resulting 3D solution is then basically the 1D solution multiplied by some prefactors. 
Our peak-height formula will be useful in determining the peak density of the 
condensate by evaluating those prefactors. 

Finally, our peak-height formula provided insights in relating all breathers generated by DT.
For example, it can be used to prove the uniqueness of the Pythagorean triplet of rogue waves.
It also provides a simple check on the accuracy of any numerical solution of
the NLS equation. This is especially useful when solving extended NLS equations with complex higher-order terms.

\begin{acknowledgments}
This research is supported by the Qatar National Research Fund (NPRPs 5-674-1-114 and 6-021-1-005), a member of the Qatar Foundation. 
S.N.N. acknowledges support from the Serbian  MESTD Grants III45016 and OI171038. M.R.B. acknowledges support by the Al-Sraiya Holding Group.
\end{acknowledgments}


\end{document}